# A Novel SLCA-UNet Architecture for Automatic MRI Brain Tumor Segmentation

Tejashwini P S[a]*, Dr. Thriveni J[a] and Venugopal K R[b]

[a]Dept. of CS&E,  University of Visvesvaraya College of Engineering, Bengaluru, India;
[b]Former Vice-chancellor,  Bangalore University, Bangalore, India.
*tejashwinirnk@gmail.com

First Author: Tejashwini P S, Research Scholar, Dept. of CS&E, University Visvesvaraya College of Engineering, Bangalore, India.
Email: * tejashwinirnk @gmail.com

# A Novel SLCA-UNet Architecture for Automatic MRI Brain Tumor Segmentation

**Abstract:** Brain tumor is deliberated as one of the severe health complications which lead to decrease in life expectancy of the individuals and is also considered as a prominent cause of mortality worldwide. Therefore, timely detection and prediction of brain tumors can be helpful to prevent death rates due to brain tumors. Biomedical image analysis is a widely known solution to diagnose brain tumor. Although MRI is the current standard method for imaging tumors, its clinical usefulness is constrained by the requirement of manual segmentation which is time-consuming. Deep learning-based approaches have emerged as a promising solution to develop automated biomedical image exploration tools and the UNet architecture is commonly used for segmentation. However, the traditional UNet has limitations in terms of complexity, training, accuracy, and contextual information processing. As a result, the modified UNet architecture, which incorporates residual dense blocks, layered attention, and channel attention modules, in addition to stacked convolution, can effectively capture both coarse and fine feature information. The proposed SLCA UNet approach achieves good performance on the freely accessible Brain Tumor Segmentation (BraTS) dataset, with an average performance of 0.845, 0.845, 0.999, and 8.1 in terms of Dice, Sensitivity, Specificity, and Hausdorff95 for BraTS 2020 dataset, respectively.

**Introduction**

During last two decades, we have noticed a tremendous augmentation in various types of diseases such as cardiovascular, chronic kidney diseases, diabetes, respiratory diseases, brain disorders, tumors and cancers. Cancer is considered as the one of the deadliest disease and one of the rapidly increasing cause of mortality worldwide. Similarly, brain tumor is the utmost noticeable sort of life threatening cancer which is also considered as the key cause for death in man and women globally (Chahal et al. 2020). Brain is the vital and paramount organ of human body which is enclosed in the skull. The brain is responsible for several functionalities such as regulating the memory, sensitive motor functions. The complete brain structure is comprised of three main components known as cerebrum, cerebellum and brain stem.

The cerebellum, which is connected to both the left and right hemispheres of the brain, is controller of carrying out higher sensory and motor activities. It is also connected to the cerebrum, which is in responsible for posture, coordinated contraction of muscles, and functional muscle movement. With the aid of the brain stem, the cerebrum and cerebellum are both attached to the spinal cord. However, the regular functions of brain are impacted by the atypical bulge of brain cells and structural problems with the brain. Despite being the deadliest, the specific reasons are still unclear, but several risk factors, such as family history and radiation exposure, have been found that might increase the chance of brain tumours.

There has been a significant surge in the incidence of brain tumors. In 2016, Nation Brain Tumor Society presented a statistical study which reported the around 78,000 individuals are identified from brain tumor and 16,500 individuals expire from brain cancer in USA every year (Mukthar et al. 2020). According to a study revealed by the International Agency for Research on Cancer (IARC), it has been found that around 126,000 individuals are diagnosed with brain tumors annually, leading to approximately 97,000 deaths (Ferlay et al. 2015). Based on the most recent population registration figures for India, it is estimated that approximately 800,000 patients succumb to cancer each year, positioning it as the nation's second-largest chronic condition (Ali et al. 2011). Therefore, early detection of brain tumor becomes the prime concern in this field of biomedical research.

Recent technological advancements have opened the door for several biomedical applications. For example, the biomedical image analysis is widely adopted in various applications such as brain tumor, kidney abnormalities, eye related disease etc. The technological advancement has led to improve the efficiency of medical imaging devices which are involved in various research and clinical treatments such as medical robots, computer-aided diagnosis and image analysis (Liu et al. 2023). The medical image analysis is helpful in guiding the medical professionals to comprehend the disease and improve the health-care quality. Thus, it is widely adopted in bran image analysis where detection of brain region, identifying the tumor and classification of tumor are the important tasks (Dong et al. 2017).

Brain tumor segmentation is a process of identifying and demarcating the boundaries of a tumor or tumors. This task can be accomplished by biomedical image analysis. Generally, MRI is considered as a promising technique in this domain. MRI images facilitate several imaging modalities such as T1-weighted (T1), contrast-enhanced T1-weighted (T1c), T2-weighted (T2), and Fluid Attenuated Inversion Recovery (FLAIR) images which helps to acquire the rich information about the brain structure and pathology. This process plays essential role in identifying and treatment of brain tumors, as it provides crucial information about the size, location and shape of the tumor(s) (Işın et al. 2016). The segmentation process typically involves the use of advanced computer algorithms that can analyze the MRI images and identify the areas of the brain that are affected by the tumor. These algorithms use various techniques, such as thresholding (Sivakumar, V et al. 2020), edge detection and clustering, to distinguish between different tissues and structures within the brain (Ramesh K. K. D et al. 2021).

Brain tumor segmentation poses several challenges, including the variability in tumor profiles and sizes, the occurrence of artifacts and noise in MRI pictures and the similarity between tumor tissues and healthy brain tissues. To overcome these challenges, researchers have developed various segmentation techniques, including manual (McGrath et al. 2020), semi-automated and fully automated segmentation [Nikan et al. 2020, Langan et al. 2022]. Manual segmentation implicates a human expert manually outlining the tumours on each MRI slice, which can be time-consuming and prone to inter-observer variability. Semi-automated segmentation techniques use computer algorithms to assist human experts in the segmentation process, while fully automated segmentation techniques rely entirely on computer algorithms. Recent developments in deep learning, predominantly with convolutional neural networks (CNNs), have led to substantial improvements in automated brain tumor segmentation. These algorithms have the ability to learn and recognize the specific features and patterns associated with brain tumors by analysing extensive datasets of MRI images. They can efficiently perform accurate segmentations, significantly reducing the time required compared to a human expert.

The current advancements in deep learning methods show great potential in CAD for medical data analysis. The rapid growths of CNN based architectures have surpassed the human level performance in several applications. Deep Learning (DL) based approaches are broadly adopted for brain tumor segmentation. The state-of-art models show the importance of encoder –decoder based architecture where UNet is considered as popular medical image segmentation method (Ronneberger et al. 2015). Recently, several UNet like architectures are introduced for medical image segmentation such as in (Myronenko A. et al. 2019) authors introduced modified UNet where an autoencoder branch is added to perform the regularization. (Jiang et al. 2020) presented cascaded UNet where a pipeline structure is introduced to segment the brain tumors. Similarly, nnUNet architecture is introduced in (Cinar N et al. 2022) with data post-processing, region based training and data augmentations. Moreover, several variants of UNet are discussed in literature review section.

The primary contributions of this research article can be summarized as follows:

- The proposed model is based on the UNet structure because it has been widely adopted mechanism to achieve better segmentation accuracy.
- The developed architecture uses dense residual blocks by replacing the residual blocks to obtain the most robust features.
- The proposed model introduces stacked convolution model.
- Further, layered and group channel attention mechanism is also incorporated.

Rest of the manuscript consists of following sections: section II describe the most recent and popular techniques of automated brain tumor segmentation, section III presents the proposed solution to improve the segmentation accuracy, section IV presents outcome of proposed approach its comparative analysis with state-of-art approaches, finally, section V presents the concluding remarks.

**Literature Survey**

(Cinar N et al. 2022) presented deep learning based segmentation technique to segment the brain tumors. Therefore, this approach uses UNet as the base architecture and incorporated pre-trained DenseNet121 to produce hybrid architecture of segmentation. This model concentrates on smaller sub-regions on tumors that upsurges the complexity of brain tumor.

(Raza et al. 2023) familiarised an end-to-end module for automated 3D BTS by using deep learning concept. This approach is obtained by combining the deep residual network and UNet architectures. As mentioned before, the traditional UNet have encoder-decoder structure to obtain the segmentation output but it suffers from the issue of disappearing gradient. Therefore, residual network is incorporated to handle this issue. Furthermore, this architecture takes into explanation both low-level and high-level attributes to make accurate predictions. In order to preserve the low-level information, it uses shortcut connection. Later, residual and convolution blocks are connected with the help of skip connections.

(Cao et al. 2022) reported that the traditional UNet performs several convolutions and pooling operations due to which spatial and contextual information is affected resulting in affecting the segmentation accuracy. In order to overcome this issue, authors introduced context attention module which considers "multiscale contextual attention information". This study utilizes an attention mechanism to effectively filter high-level attributes by incorporating spatial context information, thereby preserving important contextual details. Additionally, a channel attention model is employed to integrate both high and low-level attributes, enhancing the overall feature representation.

(Ullah et al. 2022) developed a novel fully automated procedure to segment the brain tumors. This mechanism is based on traditional UNet and employs multiscale residual attention blocks. In order to preserve the consecutive facts, this architecture uses three consecutive slices and performs multiscale learning in a cascade manner. This cascading process helps to obtain the enhanced segmented region.

(Ilhan et al. 2022) introduced an efficient segmentation approach for tumor localization and augmentation with the help of UNet architecture. In first stage, histogram based non-parametric tumor localization method is used to obtain the tumor regions. This mechanism offers significant benefits in enhancing the visual appearance of low-contrast tumors, improving their visibility and aiding in their accurate detection. The obtained enhanced tumor image is further processed with the UNet to achieve the final segmentation. Deep skip connections paired with an effective and lightweight UNet were first developed by (Deng et al. [2022]). The skip connections make it easier to completely extract the encoder module's functionalities. A novel loss function, denoted as DFK, is also developed to augment the accuracy fir ET, WT, CT.

In order to interpret the 3D brain MRI data, (Nodirov et al. 2022) developed a 3D UNet model that combines the pre-trained 3D MobileNetV2 and attention module with many skip connections. Faster convergence is made possible by MobileNet V2 while maintaining an operable model size. The extra skip connections facilitate the interchange of features across blocks in a similar manner.

Dual graph reasoning unit segmentation technique was introduced by (Ma et al. 2022) for brain tumor segmentation. This paradigm is made up of two concurrent graph reasoning modules for channel and spatial reasoning. While the graph attention network is utilised to describe the contextual interdependencies across channels, the spatial reasoning modules assist in modelling the long-range spatial dependencies.

The structural complexity and class imbalance in brain tumour segmentation were addressed by (Huang et al. 2022). The authors merged the UNet architecture with the residual learning technique in order to accomplish this. The residual block improves feature extraction and assists in streamlining training. The classifiers are trained in the following step, where batch normalisation and maximum pooling techniques are used to strengthen the regularisation procedure.

According to (Cao et al. 2022), convolution procedures prevent the conventional UNet technique from learning the global semantic information pattern. The authors developed a UNet-based transformer for medical image segmentation to address this problem. In this procedure, the image is converted into several tokens that are then fed into the U-shaped deep learning design based on the transformer and composed of skip-connection for local and global semantic feature learning. To extract the robust features, this method employs a hierarchical swin transformer technique. In a similar vein, upsampling procedures to reinstate the spatial resolution of feature maps are also carried out using patch expanding layers. Category directed attention UNet is a brand-new deep learning-based segmentation technique is introduced by (Li et al. 2022).

A Supervised Attention Module is also included in this design to capture long-term feature reliance without raising computation costs. The feature maps are also being rebuilt using an intra-class updating technique. (L Chen et al. 2018) reported that location, structure, and shape of brain tumor affect the segmentation accuracy. Therefore, a 3D CNN architecture is presented which captures the contextual information. Moreover, it uses densely linked convolution modules to boost the performance.

(Ghaffari et al. 2022) introduced 3D dense UNet architecture which uses ensemble learning model to increase the performance. Guan et al. [36] developed a deep learning architecture known as AGSE-VNet. This architecture uses SE blocks which are added to each encoder.

(Liu et al. 2022) reported that the accuracy of segmentation methods is affected due to blurred boundaries of TC and ET. Therefore, authors introduced MetricUnet, a deep learning architecture which considers voxel-level attributes to obtain the segmentation.

(Sun et al. 2021) developed 3D FCN to obtain the segmentation. This architecture uses multi-pathway arrangement of layers to extract the features from multi-modal MRI images. This architecture incorporates the use of 3D dilated convolution operations in each pathway, enabling the extraction of features from different receptive fields.

(Sheng et al. 2021) focused on extracting the second-order statistical feature of brain MRI. In this approach, the traditional slip connections are replaced by the second order modules resulting in increasing the non-linearity of segmentation network.

(Wang et al. 2021) discussed the importance of transformer mechanism which takes advantages of long-range information modelling by employing the self-attention method to increase the accuracy. Based on this concept, authors presented a transformer based 3D CNN method to segment the brain tumor. This model also follows the encoder and decoder based arrangement where encoder module uses 3D CNN to obtain the volumetric spatial feature maps whereas the decoder module leverages these attributes and employs upsampling to obtain the detailed segmentation map.

(Li et al. 2022) presented a comparative study where 3D UNet and 3D Unet ++ deep learning models are simulated to realize the segmentation performance. This study divides the process into following phases: first phase demonstrates the decomposition of tumor as WT, TC and ET. In the second phase, the deep learning models undergo training using various axial images as input. Finally, the obtained outcomes are fused to generate the final segmentation output. (Liu et al. 2023) reported that the glioma segmentation is affected due to small intensity variations between adjacent glioma areas. In order to overcome this, authors introduced attention based method. This approach uses image enhancement method to mitigate the interferences in the image background. In next stage, 3D Unet with attention layers is presented to extract the features while considering the fact of intensity dissimilarity. The attention layers help to obtain the robust attributes.

**Proposed Stacked Convolution Layered Channel Attention UNet (SLCA UNet)**

Aforementioned studies have reported the significance of deep learning based solution for biomedical image segmentation. Therefore, researchers have developed encoder-decoder based architecture called as UNet. The UNet model location information and contextual information during segmentation and combine them to obtain the segmentation map. However, higher training time, high GPU usage and scarcity of pre-trained models are some of the widely known challenges faced in UNet segmentation. The proposed worked focus on these challenges and introduce UNet based architecture for brain tumor segmentation. An overview of the conventional UNet architecture and its latest enhancements, including the integration of residual blocks was discussed. Subsequently, we delve into a concise examination of the different layers. Lastly, the SLCAUNet architecture was introduced which propose as a segmentation solution.

*UNet Segmentation*

(Ronneberger et al. 2015) introduced the UNet architecture for biomedical image segmentation, aiming to enhance the robustness of the segmentation method. The UNet architecture consists of two primary paths known as the contraction path and the expanding path.

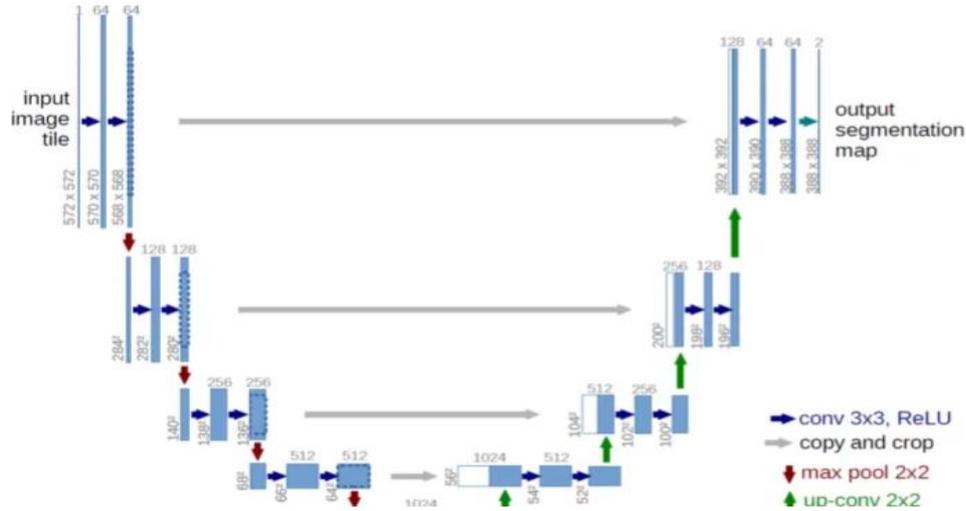

Figure 2. UNet Architecture [13]

The contraction path is also known as the encoder path. This encoder module is useful in capturing the contextual information of image. This path is constructed with the aid of convolution and max pooling operations. The second path is also known as the decoder. The decoder module is a symmetric expanding path which is used in localization with the help of transposed convolutions. Figure 2 depicts the general architecture of UNet. More details of UNet can be obtained from (Ronneberger et al. 2015).

*Residual block & Dense Block*

The convolution network can be trained to be significantly deeper, more accurate, and more efficient by reducing the distance between input and output by shortening the connections. Generally, the training loss is a hard to navigate the loss can increase due to increased number of deep layers. In order to overcome this issue, researchers have suggested incorporating cross connection between layers of network which allows large section to be skipped. This process helps to reduce the loss, and train the network with optimal weights. Below given Figure 3 (a) shows the basic representation of Residual blocks. Similarly, Figure 3 (b) shows the updated UNet architecture obtained by adding the residual block in decoder phase.

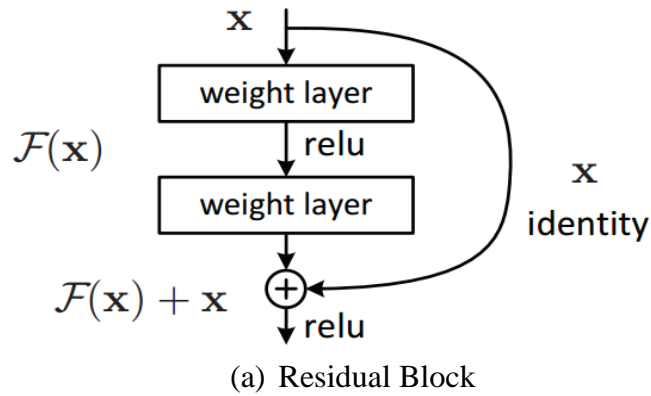

(a) Residual Block

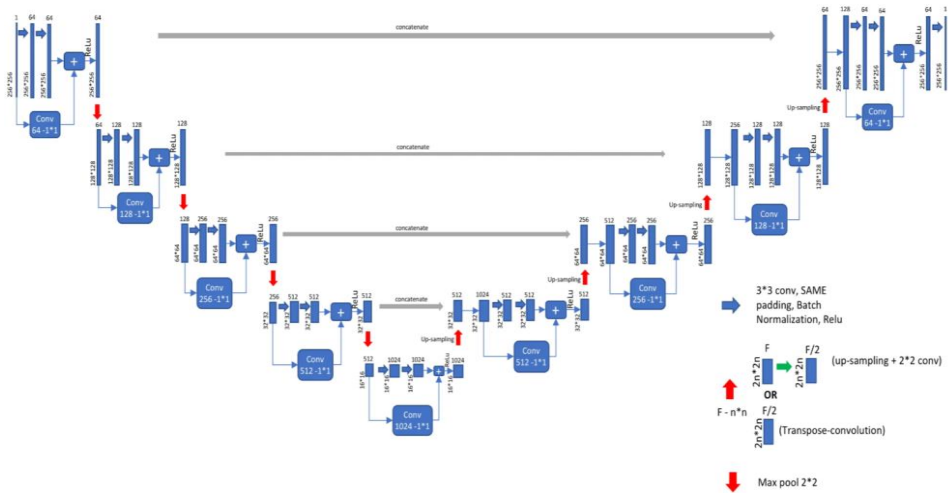

(b) Residual block combined with UNet Architecture

Figure 3. UNet architecture and residual blocks

The Residual block has two connections where first connection travels from input, convolutions, batch normalization and linear functions. On the other hand, the other connection uses skip connection over the series of convolutions. The outcome of these two block is combined together which is processed further to obtain the final map. In order to obtain the Dense block, the tensor addition can be converted into tensor concatenation. With each skip connection, the network become denser resulting in producing the DenseNet.

Incorporating these blocks leads to improve the performance of UNet. Some of the improvements in UNet architecture are listed below:
- The additional convolution operations use 3*3 filter and SAME padding therefore the feature size remains unchanged on each level of encoder and decoder path.
- The SAME padding helps to preserve the boundary information thus it allows adding more convolution operations.
- In order to concatenate the encoder feature with expanding path, it doesn't require cropping of feature map. Avoiding the cropping operation helps to reduce the information loss.
- The short skip connection which is considered as local skip connection helps to obtain the smooth loss curve and it avoids the gradient vanishing problem.

*Proposed Stacked Convolution Layered Channel Attention UNet*

This section presents the proposed SLCA UNet architecture for brain tumor segmentation task. Below given Figure 4 depicts the architectural representation of proposed SLCA UNet approach.

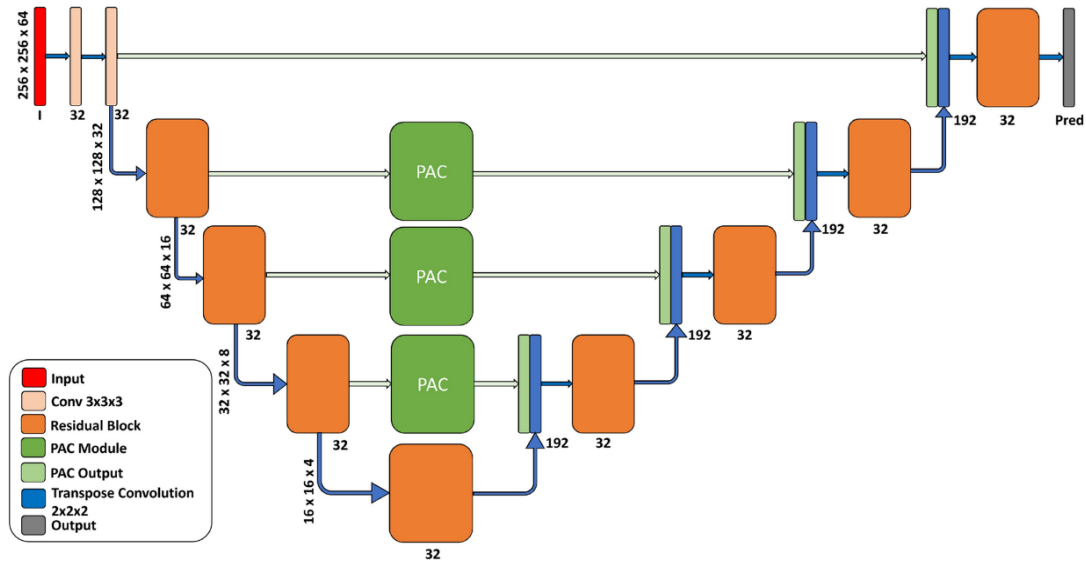

Figure 4. Architecture of proposed SLCAUNet model

The main aim of this architecture is to overcome the issues of UNet. Therefore, the proposed architecture consists of Layered attention and Channel attention mechanism as backbone of the network along with this, the proposed architecture focus on minimizing the parameter count which improves the network information and gradient flow.

To address the challenges of feature redundancy and the loss of edge information in the feature map, a feature generation process was introduced. This process involves producing features of different scales and passing them to the decoder side. The proposed solution includes a stacked convolution module placed over the skip-connection, effectively mitigating the aforementioned issues.

Similarly, the conventional convolution blocks have been replaced with residual dense blocks to facilitate the downsampling of feature maps through the utilization of strided convolutions. Below given Figure 5 (a) shows the architecture of residual dense block and Figure 5 (b) shows the architecture of stacked convolution module.

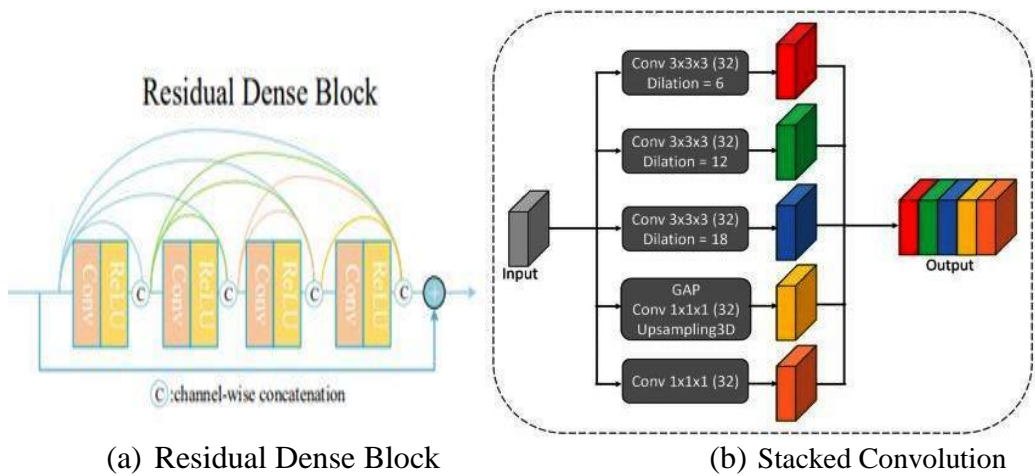

(a) Residual Dense Block          (b) Stacked Convolution

The convolution operations in this module are mathematically defined as follows:
$$Conv_{m \times m \times m}(x, s, K; \theta) = f(w^j \otimes_s x + b^j), \forall 1 \leq j \leq K, w^j \in \theta, b^j \in \theta$$
Where $x$ denotes the input feature map, $s$ denotes the convolution's stride, $K$ represents the number of kernels, $m$ characterises the kernel dimension, $\theta$ represents the weight and biases of the considered kernels, $f$ is the activation function, $\otimes_s$ denotes the convolution task with stride, $w^j$ is weight and $b^j$ is the bias of $j^{th}$ kernel. Based on these parameters the operations of residual block can be explained as:
$$c_1^i = Conv_{m \times m \times m}(c^{i-1}, s_0, K; \theta_1^i)$$
$$c_2^i = Conv_{m \times m \times m}(c^{i-1}, s_0, K; \theta_2^i)$$
$$c_3^i = Conv_{m \times m \times m}(c^{i-1}, s_0, K; \theta_3^i)$$
$$c^i = c_2^i \oplus c_3^i,$$
Where $c^{i-1}$ and $c^i$ represents the input and outputs, respectively.

$c_1^i, c_2^i$ and $c_3^i$ denotes the output of convolution operations and $\oplus$ represents the element-wise addition.

Additionally, the proposed solution introduced layered attention and channel attention modules to enhance the accuracy of segmentation. Figure 6 depicts the integrated architecture of the layered and channel attention mechanisms. The layered module enables the extraction of weights for high-level, low-level and detailed features. Meanwhile, the channel attention block utilizes squeeze-and-excitation (SE) blocks to process the weights of channels for improved learning. This helps to obtain the multiscale contextual information which can be expressed as:
$$M_i^A = M_{i*r_1} K_1^i + M_{i*r_2} K_2^i, i = 1,2,3, ....$$
Where $M_1, M_2$ and $M_3$ represents the high, low and detailed features, respectively. $K_1^i$ and $K_2^i$ represents the first and second atrous convolution operation with $r_1$ and $r_2$ dilation rates.

In order to obtain the global distribution of feature maps, perform global average pooling operation which is expressed as:
$$I_j = \frac{1}{X \times Y} \times \sum_x \sum_y M(x, y, j) \quad j = 1, 2, ... K$$

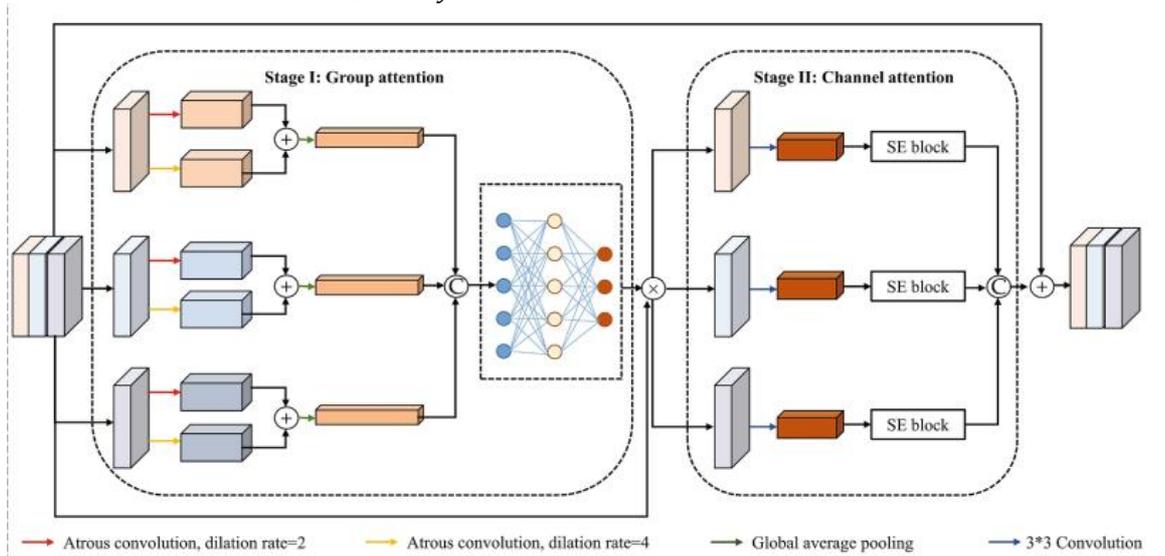

(c)

Figure 1. Layered and channel attention module.

In next stage, a dense neural network that helps to generate the weights of groups was introduced. The output of first and second dense layer can be expressed as:

$$O = f_2(w_2^T \times f_1(W_1^T \times I + B_1) + B_2)$$

$$G_i = \frac{\exp \exp [O(i)]}{\sum_{j=1}^{3} \exp \exp [O(j)]}, i = 1,2,3$$

Where $W$ symbolizes the weights and $B$ symbolizes the biases of dense layers, $f_1$ & $f_2$ denotes the activation function and $G$ signifies the group weight. In this study, the rectified linear activation function has been utilized. After acquiring the weighted groups, three squeeze-and-excitation (SE) blocks are employed to build the weights for each channel within each feature group. Below given Figure 7 illustrates the architecture of this module.

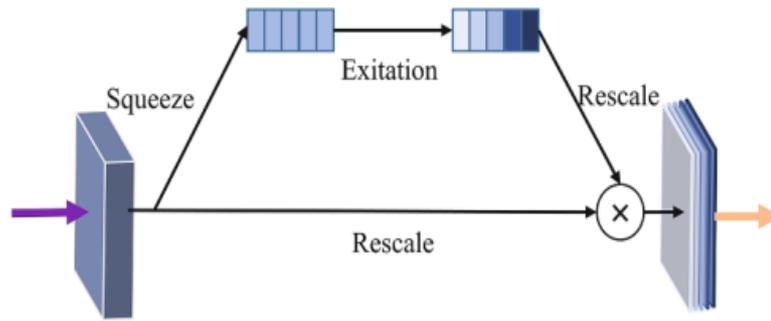

Figure 7. Architecture of SE block

The global average pooling operation is employed to this SE block which is expressed as:

$$C_j^i = \frac{1}{X \times Y} \sum_x \sum_y M_R^i(x, y, j), \ j = 1,2, \dots K$$

This generates the final feature amp which is processed on the decoder side until the final segmented map is obtained by applying the decoder operations.

**Results and Discussion**

In this segment, the outcome of proposed SLCA UNet architecture is presented and obtained output is matched with state-of-art segmentation methods. Initially, we present a brief information about dataset used in this work. Later, performance evaluation parameters are discussed.

*Dataset Details*

To assess and verify the performance of the SLCAUNet model, three benchmark datasets from the Multimodal Brain Tumor Segmentation Challenge (BraTS) 2017, 2018, and 2019 were employed.

*BraTS 2017 Dataset*

The dataset utilized in this study consists of 285 patients who have been diagnosed with Glioma, a type of brain tumor. Among these patients, 210 were diagnosed with high-grade glioma (HGG), while 75 were diagnosed with low-grade glioma (LGG). The dataset includes four different modalities such as Flair, T1, T1ce, and T2 for each case. The data has been manually categorized into four distinct groups, which include healthy tissue, necrosis and non-enhancing regions, edema and enhancing tumor. Additionally, there is a validation set comprising 46 patients with an unknown grade, enabling further evaluation of the model's performance.

### *BraTS 2018 & 2019 Dataset*

It contains training set same as used in BraTS 2017 dataset but it has varied validation set where 66 unlabelled patients' data are considered. The BraTS 2019 dataset contains more number of samples than BraTS 2017 and 2018. Specifically, the training set encloses total 335 glioma patients where 259 sample belong to HGG case and 76 for LGG cases. Moreover, the number of validation patient data is also increased to 125.

### *Performance Measurement Parameters*

In order to assess the performance, several performance measurement parameters such as Dice Similarity Coefficient (DSC) and Hausdroff distance were employed. The Dice Score is used to measure the overlapped area between actual labelled region and predicted region. The dice coefficient can be measured as follows:

$$DSC = \frac{2T^P}{F^N + F^P + 2T^P}$$

Where $T^P, F^P$, and $F^N$ denotes the true positive, false positive, and false negative values.

On the other hand, the Hausdroff distance can be estimated as follows:

$$HD(T,P) = \{d(t,p), d(t,p)\}$$

Where $sup$ represents the supremum and $inf$ is used to denote the infimum, $t$ denotes the points on surface $T$ of groundtruth, and $p$ denotes the points on the surface $P$ of the predicted region, $d(.,.)$ is the distance between points $t$ and $p$.

### *Comparative analysis*

This segment explains the comparative investigation of proposed SLCAUNet for aforementioned publicly available datasets where SLCAUNet approach is implemented on training and validation datasets. First of all, we present the outcome of SLCAUNet approach for BraTS 2017 dataset where entire 285 training images are considered for experiment out of which 171 images are used as training set, 57 images are used for testing set and 57 images are used for testing set. Below given table shows the comparative analysis for the considered dataset.

Table .1. Dice Score Performance for Whole Tumor, Core Tumor, and Enhancing Tumor.

| Method | Whole | Core | Enhancing |
|---|---|---|---|
| UNet | 0.87 | 0.763 | 0.704 |
| ResUNet | 0.874 | 0.767 | 0.715 |
| AGU Net | 0.875 | 0.773 | 0.712 |
| AGResUNet | 0.76 | 0.78 | 0.72 |
| **Proposed Model** | **0.885** | **0.813** | **0.805** |

The performance of SLCAUNet model is measured for three scenarios of brain tumor such as Whole, core and enhancing tumor. The obtained performance is compared with state-of-art algorithms as mentioned in (Zhang et al. 2020). The SLCAUNet approach obtained the average DSC as 0.885, 0.813, and 0.805 for WT, CT and ET tumor. Similarly, we measured the performance of SLCAUNet approach for HGG. BraTS 2017 has 210 HGG cases which are further divided into two parts as 168 for training and 42 cases for testing set. Below given table shows the obtained performance for HGG cases in BraTS 2017 dataset.

Table. 2. Dice score comparisons for HGG samples in BraTS 2017 dataset

| Method/Author | Whole | Core | Enhancing |
|---|---|---|---|
| UNet | 0.881 | 0.847 | 0.814 |
| ResUNet | 0.886 | 0.857 | 0.823 |
| (Chen et al. 2018) | 0.72 | 0.847 | 0.81 |
| (Kamnitsas et al. 2017) | 0.9 | 0.857 | 0.730 |
| (Dong et al. 2017) | 0.831 | 0.75 | 0.75 |
| (Pereira et al. 2016) | 0.84 | 0.801 | 0.62 |
| (Kermi et al. 2018) | 0.88 | 0.72 | 0.82 |
| (Zhao et al. 2018) | 0.865 | 0.85 | 0.816 |
| AGResUNet (Zhang et al. 2020) | 0.891 | 0.865 | 0.830 |
| (Zhang et al. 2022) | 0.891 | 0.865 | 0.83 |
| (Zhao et al. 2018) | 0.9 | 0.83 | 0.78 |
| (Guan eta al. 2022) | 0.69 | 0.85 | 0.68 |
| *Proposed Model* | *0.921* | *0.895* | *0.887* |

According to this investigation, the suggested DL based model attains average DSC performance as 0.921, 0.895, and 0.887 for Whole, Core and Enhancing tumor cases respectively. The performance evaluation of the SLCAUNet model for the validation dataset, which includes a total of 285 MRI images used for training. The obtained performance results for the validation dataset are presented in Table 3. In this experiment, DSC and Hausdorff 95 measurement parameters are used.

Table. 3. Comparative analysis for BraTS 2017 validation dataset

| Author | DSC | | | Hausdorff 95 | | |
|---|---|---|---|---|---|---|
| | Whole | Core | Enhancing | Whole | Core | Ehancing |
| (Islam et al. 2017) | 0.875 | 0.75 | 0.688 | 9.81 | 12.2 | 12.93 |
| (Jesson et al. 2017) | 0.898 | 0.752 | 0.712 | 4.15 | 8.64 | 6.97 |
| (Kamnitsas et al. 2017) | 0.902 | 0.796 | 0.737 | 4.22 | 6.55 | 4.501 |
| (Pereira et al. 2018) | 0.883 | 0.772 | 0.718 | 6.201 | 10.23 | 6.71 |
| (Hu et al. 2017) | 0.850 | 0.70 | 0.650 | 25.24 | 21.45 | 17.98 |
| AGResUNet (Zhang et al. 2020) | 0.88 | 0.781 | 0.749 | 6.87 | 9.32 | 3.74 |
| *Proposed Model* | *0.915* | *0.852* | *0.831* | *3.80* | *5.10* | *3.55* |

The experimental analysis conducted on the BraTS 2017 training and validation datasets, which demonstrates a major improvement in the performance of the SLCAUNet model.

Furthermore, the comparative analysis of the SLCAUNet model with other methods was carried out on the BraTS 2018 validation dataset, in terms of DSC and Hausdorff metrics. Table 4 presents the comparative analysis results for whole, core, and enhancing tumors.

Table. 4. Comparative analysis for BraTS 2018 validation dataset

| Method/Author | DSC | | | Hausdorff 95 | | |
|---|---|---|---|---|---|---|
| | Whole | Core | Enhancing | Whole | Core | Enhancing |
| UNet (Zhang et al. 2020) | 0.867 | 0.793 | 0.751 | - | - | - |
| ResUNet (Zhang et al. 2020) | 0.870 | 0.808 | 0.760 | - | - | - |
| AGUNet (Zhang et al. 2020) | 0.871 | 0.798 | 0.771 | - | - | - |
| (Cicek etal. et al. 2016) | 0.885 | 0.718 | 0.760 | 17.10 | 1160 | 6.04 |
| (W Chen et al. 2018) | 0.894 | 0.831 | 0.749 | - | - | - |
| (N Nuechterlein et al. 2018) | 0.883 | 0.814 | 0.737 | - | - | - |
| (G Wang et al. 2018) | 0.873 | 0.783 | 0.751 | 5.90 | 8.03 | 4.53 |
| (S Chandra et al. 2018) | 0.872 | 0.795 | 0.741 | 5.04 | 9.59 | 5.58 |
| (K Hu et al. 2019) | 0.882 | 0.748 | 0.718 | 12.60 | 9.62 | 5.69 |
| AG-ResUNet (Zhang et al. 2020) | 0.872 | 0.808 | 0.772 | - | - | - |
| *Proposed Model* | *0.915* | *0.852* | *0.831* | *3.80* | *5.10* | *3.55* |

Similarly, the experimental analysis is conducted for BraTS 2019 and 2020 validation dataset. Below given table 5 show the comparative analysis for BraTS 2019 dataset.

Table. 5. Comparative analysis for BraTS 2019 validation dataset

| Method | Dice | | | Sensitivity | | | Specificity | | | Hausdorff 95 | | |
|---|---|---|---|---|---|---|---|---|---|---|---|---|
| | TE | TW | TC | TE | TW | TC | TE | TW | TC | TE | TW | TC |
| S2MetricUnet (Cao et al. 2022) | 0.84 | 0.83 | 0.72 | 0.76 | 0.86 | 0.73 | | | | | | |
| 3D FCN (Li et al. 2022) | 0.76 | 0.89 | 0.78 | | | | | | | | | |
| ResUNet (Zhang et al. 2020) | 0.72 | 0.87 | 0.78 | | | | | | | 5.97 | 9.35 | 11.47 |
| TransBTS (L. Chen et al. 2018) | 0.78 | 0.88 | 0.81 | | | | | | | 5.90 | 7.59 | 7.58 |
| Cascade 3D UNet (K. Kamnitsas et al. 2017) | 0.80 | 0.86 | 0.83 | 0.84 | 0.90 | 0.83 | | | | 6.14 | 4.92 | 6.75 |
| AMMGS (H. Dong et al. 2017) | 0.76 | 0.89 | 0.81 | 0.82 | 0.94 | 0.85 | 0.99 | 0.98 | 0.99 | 5.17 | 8.21 | 7.23 |
| *Proposed Model* | *0.84* | *0.92* | *0.86* | *0.88* | *0.94* | *0.89* | *0.99* | *0.99* | *0.99* | *4.80* | *4.5* | *6.5* |

Similarly, we extend the experimental analysis and measured the performance on BraTS 2020 dataset in terms of dice, sensitivity, specificity, and Hausdorff95. The obtained performance is demonstrated in table 6.

Table .6. Comparative analysis for BraTS 2020 dataset.

| Method | Dice | | | Sensitivity | | | Specificity | | | Hausdorff 95 | | |
|---|---|---|---|---|---|---|---|---|---|---|---|---|
| | TE | TW | TC | TE | TW | TC | TE | TW | TC | TE | TW | TC |
| MIT AU (A. Kermi et al. 2018) | 0.57 | 0.73 | 0.61 | 0.52 | 0.77 | 0.62 | 0.99 | 0.99 | 0.99 | 38.87 | 20.81 | 24.2 |
| Prob UNet (X. Zhao et al. 2018) | 0.68 | 0.81 | 0.71 | 0.69 | 0.84 | 0.69 | 0.99 | 0.99 | 0.99 | 46.88 | 41.52 | 26.2 |
| 3DDMFNet (Zhao et al. 2022) | 0.74 | 0.87 | 0.74 | 0.75 | 0.87 | 0.71 | 0.99 | 0.99 | 0.99 | 3.92 | 9.42 | 10.0 |
| AHM3D | 0.71 | 0.88 | 0.74 | 0.74 | 0.92 | 0.74 | 0.99 | 0.99 | 0.99 | 38.31 | 6.88 | 32 |
| (Ghaffari et al. 2022) | | | | | | | | | | | | |
| DS3D UNet (H. Dong et al. 2017) | 0.78 | 0.88 | 0.81 | 0.79 | 0.91 | 0.78 | 0.99 | 0.99 | 0.99 | 23.86 | 7.30 | 8.16 |
| AMMGS (Liu et al. 2023) | 0.78 | 0.88 | 0.81 | 0.79 | 0.92 | 0.80 | 0.99 | 0.99 | 0.99 | 23.61 | 7.16 | 7.98 |
| ***Proposed Model*** | ***0.82*** | ***0.90*** | ***0.84*** | ***0.83*** | ***0.94*** | ***0.84*** | ***0.99*** | ***0.99*** | ***0.99*** | ***24.50*** | ***7.30*** | ***8.1*** |

According to the comparative analysis presented in table 6 the proposed SLCAUNet approach achieves better performance in terms of Dice score, sensitivity, specificity, and Hausdorff95. Below given Figure 8 shows the qualitative segmentation outcome by using proposed SLCAUNet approach.

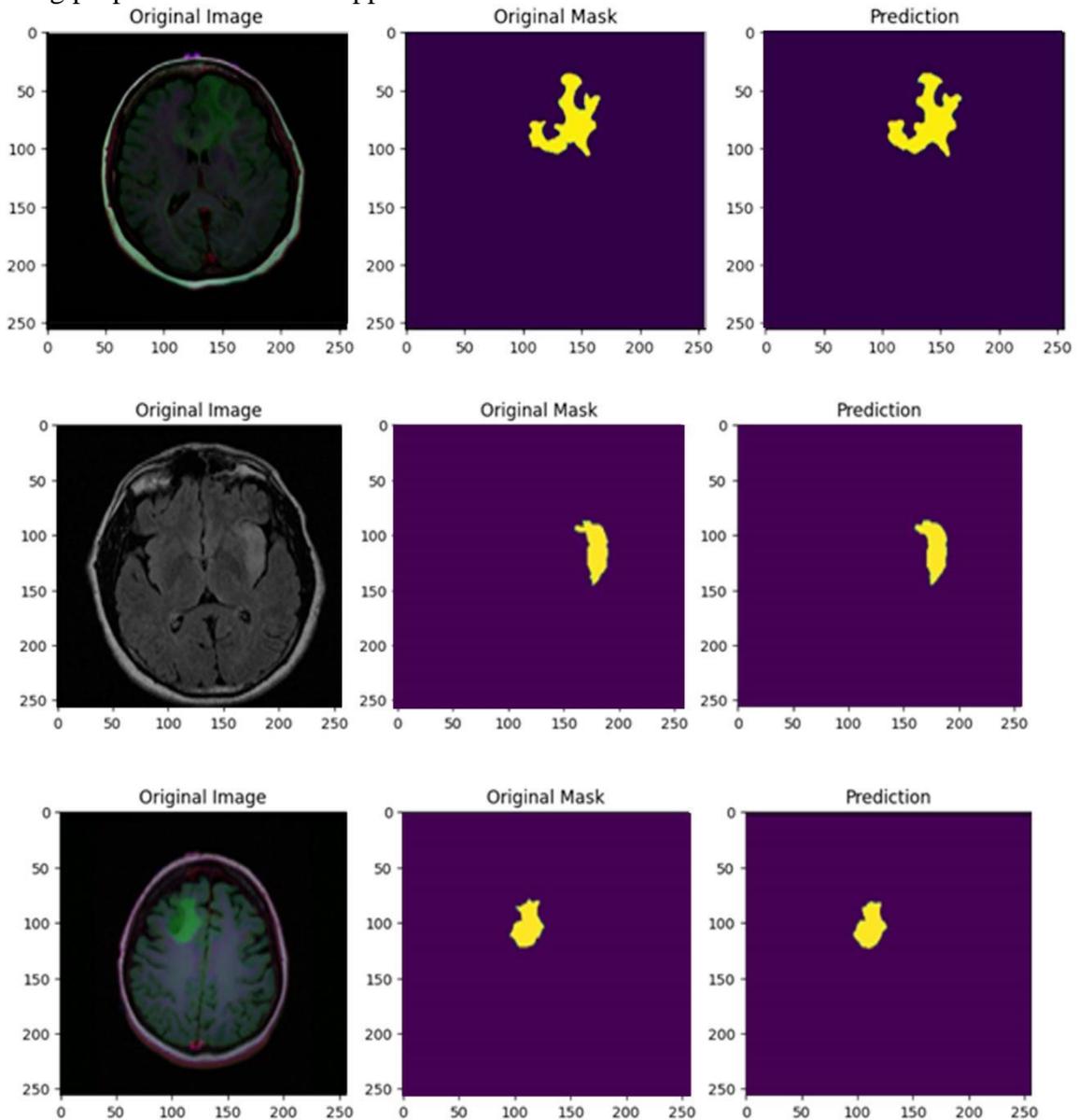

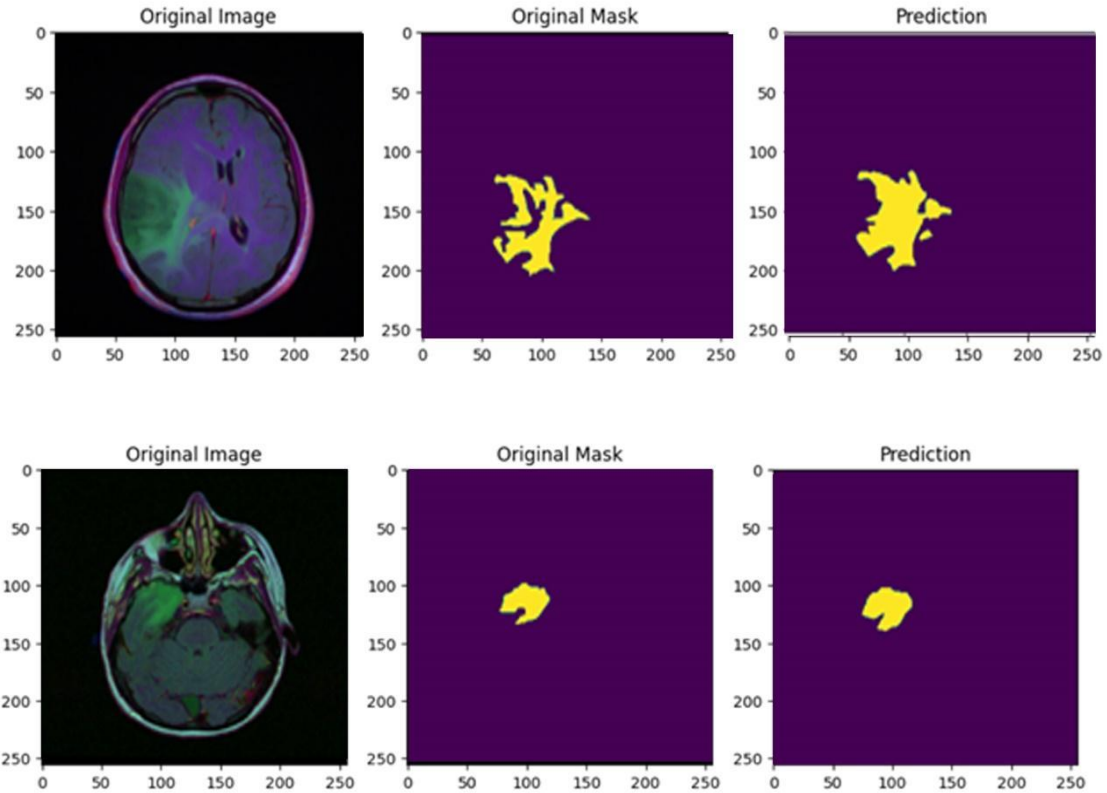

**Conclusion**

In this work, we have focused on deep learning based method for automated brain tumor segmentation. The literature review revealed the importance of UNet architecture for biomedical image analysis. Several researches have been carried out based on this module. However, the traditional methods suffer from several challenges such as contextual information, training time, and poor accuracy. Therefore, we presented a novel deep learning based UNet architecture which uses UNet as base model and incorporated several changes such as traditional convolutions are replaced by dense layers, layered attention and channel attention modules are also included to obtain the robust contextual information. Moreover, we have included stacked convolution block to obtain the robust feature map. The robustness of proposed SLCAUNet model is measured on publically available datasets such as BraTS 2018, BraTS 2019 and 2020. The experimental investigation demonstrates that the proposed DL based methodology realizes improved segmentation performance in terms of dice score, specificity, sensitivity, and Hausdorff95.